\documentclass[preprint,12pt]{aastex}
\usepackage{epstopdf}
\begin{document}

\def\putplot#1#2#3#4#5#6#7{\begin{centering} \leavevmode
\vbox to#2{\rule{0pt}{#2}}
\includegraphics{#1}

\end{centering}}


\slugcomment{Submitted to ApJ}

\shorttitle{Z Cam shell expansion}
\shortauthors{Shara et al}

\title{The Inter-Eruption Timescale of Classical Novae from Expansion of the Z Camelopardalis Shell}

\author{Michael~M.~Shara\altaffilmark{1,2}, Trisha~Mizusawa\altaffilmark{1}, David~Zurek\altaffilmark{1,2}, Christopher D. Martin\altaffilmark{3}, James D. Neill\altaffilmark{3} and Mark Seibert\altaffilmark{4}}


\altaffiltext{1}{Department of Astrophysics, American Museum of Natural
History, Central Park West and 79th street, New York, NY 10024-5192}
\altaffiltext{2} {Visiting Astronomer, Kitt Peak National Observatory, National Optical Astronomy Observatory, which is operated by the Association of Universities for Research in Astronomy (AURA) under cooperative agreement with the National Science Foundation. }
\altaffiltext{3}{Department of Physics, Math and Astronomy, California Institute of Technology, 1200 East California Boulevard, Mail Code 405-47, Pasadena, California 91125}
\altaffiltext{4}{Observatories of the Carnegie Institution of Washington, 813 Santa Barbara Street, Pasadena, California 91101} 

\begin{abstract}
 
 The dwarf nova Z Camelopardalis is surrounded by the largest known classical nova shell. This shell demonstrates that at least some dwarf novae must have undergone classical nova eruptions in the past, and that at least some classical novae become dwarf novae long after their nova thermonuclear outbursts. The current size of the shell, its known distance, and the largest observed nova ejection velocity set a lower limit to the time since Z Cam's last outburst of 220 years.  The brightest part of the Z Cam shell's radius is currently p $\sim 1690$ pixels. No expansion of the radius of the brightest part of the ejecta was detected, with an upper limit of pdot $\leq 1 $pixel/3 years). This suggests that the last Z Cam eruption occurred p/pdot $\geq$ 5,000 years ago. However, including the important effect of deceleration as the ejecta sweeps up interstellar matter in its snowplow phase reduces the lower limit to 1300 years. This is the first strong test of the prediction of nova thermonuclear runaway theory that the interoutburst times of classical novae are longer than 1000 yr. The intriguing suggestion that Z Cam was a bright nova, recorded by Chinese imperial astrologers in October - November 77 BCE, is consistent with our measurements. If Z Cam was indeed the nova of 77 BCE we predict that its ejecta are currently expanding at 85 km/s, or 0.11 arcsec/yr. Detection and measurement of this rate of expansion are doable in just a few years.  

\end{abstract}


\section{Introduction and Motivation}

Dwarf and classical novae are all close binary stars, wherein a white dwarf accretes hydrogen-rich matter from a Roche-lobe filling companion or possibly the wind of a nearby giant. A classical nova results from a thermonuclear runaway (TNR) in the degenerate layer of accreted hydrogen on the white dwarf's surface. The TNR causes the rapid rise to  $\sim 10^{5} L_\sun$ or more, and the ejection of the accreted envelope (\citet{sha89} and \citet{yar05}) which has a mass Menv of order $10^{-5}M_\odot$. In dwarf novae, an instability \citep{osa74} episodically dumps much of the accretion disk onto the white dwarf. The liberation of gravitational potential energy then brightens these systems by up to 100-fold every few weeks or months \citep{war95}. "Combination novae", involving both a disk instability and a thermonuclear runaway, may also occur \citep{sok06}. 

TNR theory predicts that the white dwarfs in all dwarf novae will eventually accrete enough mass to undergo classical nova eruptions \citep{sha86}. Before 2007 none of the $\sim 500$ known dwarf novae had yet been reported to have undergone a classical nova eruption, but this is not very surprising. With accretion rates Mdot that are of order several times $10^{-10}M_\odot$/yr, one would have expected any given dwarf nova to erupt as a classical nova about once in Menv/Mdot $\sim10,000$ to 100,000 years. Hence only of order one classical nova is expected to have erupted from the known sample of dwarf novae during the past century. 

The identified progenitors of almost all classical novae are novalike variables, in which Mdot is too high to permit the disk instability that drives dwarf nova eruptions. \citet{col09} have updated the seminal work of \citet{rob75}, finding no evidence for dwarf nova eruptions in the decades before classical nova explosions. If the hibernation scenario of cataclysmic variables \citep{sha86} is correct then one expects to see dwarf novae associated with classical novae several centuries before or after a classical nova erupts. This is because during the few centuries before a nova eruption the increasingly high mass transfer rate of a novalike variable chokes off dwarf nova eruptions. During the few centuries after a nova eruption the mass transfer rate remains high due to irradiation of the red dwarf. The highest mass transfer rate dwarf novae are the Z Camelopardalis stars, and within the context of the hibernation scenario one expects these stars to be the most likely progenitors of the novalike variables before they erupt as classical novae. One also expects old novae to evolve from novalike variables to Z Cam stars in the centuries after nova eruptions. Thus the hibernation scenario predicts that some, but not all Z Cam stars should be surrounded by old nova shells.

In 2007 we reported the discovery of a nova shell surrounding the prototypical dwarf nova Z Camelopardalis \citep{sha07}. This shell is an order of magnitude more extended than those detected around any other classical nova.The derived shell mass matches that of classical novae, and is inconsistent with the mass expected from a dwarf nova wind or a planetary nebula. The Z Cam shell observationally linked, for the first time, a prototypical dwarf nova with an ancient nova eruption and the classical nova process. This was the first-ever confirmation of a key prediction of cataclysmic binary TNR theory:  the accreting white dwarfs 
in dwarf novae must eventually erupt as classical novae.

Nova shells are observed to expand measurably on timescales of years and decades \citep{due87}. Determining the angular expansion rate of a nova shell offers the opportunity to determine (for novae whose eruption was not detected) when the explosion occurred \citep{sch10}. Astronomers have only been regularly detecting Galactic novae since the 1880s. Except for 10 recurrent novae which display eruptions just decades apart \citep{sch10}, and which probably involve massive white dwarfs, all but one known Galactic classical novae have eruption recurrence times longer than 130 years. As of 2012, WY Sge (the now-recovered classical nova of 1783 A.D. \citep{sha84} has a recurrence time longer than 229 years. This is still a very weak observational limit. TNR theory predicts inter-eruption times of $\sim1000$ to 100,000 years \citep{yar05}. Since the Z Cam shell is, by far, the largest classical nova shell ever detected, it is reasonable to assume that it is also the {\it oldest} shell yet detected. Thus the Z Cam shell offers the first opportunity to place a much stronger, direct observational lower limit on the time between nova eruptions. This is also a direct test of the TNR theory of nova explosions. 

In Section 2 we describe our new observations of the Z Cam ejecta. We show the deepest imagery yet of that shell in Section 3. In section 4 we compare images from 2007 and 2010 to place limits on the expansion and possible age of the ejecta. We briefly summarize our results in Section 5.  
  
\section {Observations and Image Processing}

Narrowband images in the lines of H$\alpha$ and [NII], and broadband R images  were obtained with the Mosaic CCD camera at the prime focus of the Kitt Peak National Observatory Mayall 4 meter telescope. The Mosaic camera on the 4 meter telescope has eight 2048 x 4096 SITe thinned CCDs, and an image scale of 0.26 arcsec/pixel.  Imaging was carried out on the nights of 23 and 24 January 2007, and 07 and 09 February 2010. 10 R band images of 180 seconds each duration (1800 seconds total exposure), 61 Halpha +[NII] images of 300 seconds each, and 45 Halpha +[NII] images of 600 seconds each (a total of 45,300 seconds exposure) were obtained in 2007. In 2010, 19 R band images of 180 seconds each duration (3420 seconds total exposure) and 19 Halpha +[NII] images of 1800 seconds each (34,200 seconds exposure) were obtained. Images were dithered over both nights during each epoch. All 2007 images were binned 2x2 as the data were taken to improve the S/N. Conditions were generally clear on all four nights. 

After flatfielding and de-biassing, standalone Daophot \citep{ste87} was used to align the images on each chip; then all of the chips were matched together.  All of the continuum (hereafter "R") and all of the narrowband (hereafter "[NII]") images of each epoch were combined to create the deepest possible image.  The images were stitched together using montage2, a mosaicking program within the standalone Daophot.  After this process was completed individually for both the [NII] and R band images, the narrow and broadband images were matched up with Daophot (which uses triangular stellar patterns for its matching algorithm). The 2010 images were then binned 2x2 to match the 2007 images.

Ultraviolet imagery was also obtained with the NASA GALEX satellite. The GALEX image data include 
far-UV (FUV; $\lambda_{eff}$=1516~\AA, $\Delta\lambda$=256~\AA) and near-UV 
(NUV; $\lambda_{eff}$=2267~\AA, $\Delta\lambda$=730~\AA) images in circular fields of diameter
$1\fdg2$. The total exposure in the FUV filter is 4.1 Ksec while that in the NUV filter is 11.9 Ksec.
The spatial resolution is $\sim$5". Details of the GALEX
instrument and data characteristics can be found in \citep{mar05} 
and \citep{mor05}. The imaging data have been
processed under the standard GALEX survey pipeline. Details about the
pipeline can found at $http://www.galex.caltech.edu/DATA/gr1\_docs.$

\section {Imaging of the Z Cam Shell}

The resulting H$\alpha$+[NII] and R images from 2010 are shown in Figures 1 and 2, respectively. Z Cam is circled in each image. This is one of the deepest narrowband-broadband image pairs ever taken of any nova ejecta. The narrowband image is dominated by the striking hemispherical arc to the SW, already clear in the discovery images of \citet{sha07}. The fainter linear features to the SE and NE, barely detected in \citet{sha07} are also clearly seen. Most surprising of all is the broad linear feature SW of the hemispherical arc. It is visible in the GALEX Far-UV, but only marginally so in the GALEX near-UV images. We originally suspected that this feature is due to light reflected by interstellar cirrus, but the deep narrowband image of Figure 1 demonstrates that it is 
H$\alpha$+[NII] emission-dominated.

The difference image (H$\alpha$+[NII] minus R) is shown in Figure 3. Z Cam is indicated with 4 arrows. A dozen or more faint arcs of emission, smaller and fainter than the bright hemisphere to the SW of Z Cam are apparent. The linear features located about 14 arcmin to the NE, SE and SW of Z Cam are not matched by a similar feature to the NW. The ejecta of classical novae often display geometries consistent with ejection of an equatorial ring and polar blobs. The geometry of the emission of Figure 3 suggests that the NE, SE and SW linear features represent the current boundaries of the equatorial ejecta projected on the plane of the sky. The bright SW arc of nebulosity could be where polar ejecta have been decelerated by interstellar matter, but the one-sidedness of the arc - the absence of a NE arc - is puzzling. 

In Figure 4 we have differenced part of the net emission image of Figure 3 with itself, after successively larger displacements of 1, 2 and 4 pixels in each of the x and y coordinates. Figure 4, zoomed on the SW arc, is thus the derivative of Figure 3. The edge of the brightest part of the arc stands out in relief in the 1 and 2 pixel displacement images. Fainter edges are visible in the 4 pixel displacement frame. 

A direct comparison of the Galaxy Explorer (GALEX) satellite's summed images in the Far-UV and Near-UV channels with the optical narrowband image of Figure 1 is shown in Figure 5. The morphologies of the ejecta are remarkably similar, with one notable exception. The linear feature SW of the arc is prominent in both the FUV and H$\alpha$ + [NII] images, but absent in the GALEX NUV image. The strong presence of the linear feature in the net H$\alpha$ +[NII] image of Figure 3 proves that this matter is emitting (and not just scattering) light. The SW linear feature is thus similar to the SE and NW linear features that bound Z Cam. Very faint wisps of emission nebulosity are also visible to the NW of Z Cam, but no linear feature is visible. We suggest that the reason we see the arc in GALEX FUV but not NUV light is that the emission mechanism is molecular hydrogen (H2) fluorescence \citep{mar90}. The GALEX NUV channel passband is too red to detect H2 emission. Unfortunately the GALEX grism images of Z Cam are too confused and/or do not go deep enough to determine the FUV emission spectrum. 

\section {Expansion and Age of the Z Cam Shell}

As noted in the introduction, classical nova eruption recurrence times are currently only weakly constrained by observations of old novae to be greater than about 130 years. This is much shorter than the predictions of TNR theory inter-eruption times of $\sim1000$ to 100,000 years \citep{yar05}. The size of the Z Cam ejecta and its accurately-known distance allow us to put a hard lower limit on the time since its eruption as follows. 

Virtually all classical novae (with the exception of the 10 known Galactic recurrent novae) exhibit ejection velocities in the range 300-3000 km/s (cf \citet{war95}, Table 5.2). The most extended emissions surrounding Z Cam are the linear features located 14 arcmin NE, SW and SE of the nova. At the 163 pc distance of Z Cam \citep{tho03} these features are located about r = 0.7 pc from Z Cam. Traveling at 3000 km/s, with no deceleration, the Z Cam ejecta could have reached their present size in no less than 220 yr. As we now show, the ejecta must have been significantly decelerated via interactions with the interstellar medium (ISM), so the time since Z Cam's last nova eruption is considerably greater than 220 yr. 

The ejecta of novae suffer significant deceleration on a timescale of a century, as suggested by \citet{oor46} and directly measured by \citet{due87}. The Oort "snowplough" \citep{oor51} model assumes conservation of momentum and energy, and a simple momentum exchange between the ejecta and the ISM. Velocity half-lifetimes (i.e. the times after which the ejecta expansion velocity drops to half its initial value) of 48 and 102 years, respectively, are predicted by the Oort model for ejected shells with masses of $10^{-5}M_\odot$ and $10^{-4}M_\odot$ and ejection velocities of 1,000 km/s.

The observed expansion velocity dropped to half its initial value in 65, 58, 117 and 67 years, respectively, for the four classical novae V603 Aql, GK Per, V476 Cyg and DQ Her \citep{due87}. The observed peak ejection velocities of these four novae were 1700, 1200, 725 and 325 km/s. The remarkably small range in deceleration half-lifetimes t for the four novae noted above are in excellent accord with the snowplough model's predictions. Z Cam's ejecta must have undergone at least two deceleration half-lifetimes in the more than 220 years since its eruption. The ejecta have therefore evolved from the blastwave into the momentum-conserving snowplow phase of explosions into a uniform medium \citep{cio88}.  
 
Our chief goal in obtaining the second epoch images of Z Cam's ejecta was to determine if any expansion of the shell could be seen between the two epochs. Only the SW arc is bright enough to allow us to attempt this measurement. We first tried to map the edge of the arc, the most prominent part of the shell, relative to a set of stars. The "edge" of the arc is spread over several pixels by the telescope/camera PSF. Several methods of measuring the arc in each epoch showed that our data could not convincingly demonstrate any movement in the shell.  We therefore shifted the 2010 epoch image (which, as noted in Section 2, was binned 2x2 to match the 0.52 arcsec pixels of our 2007 data) by 0.5, 1, and 2 pixels in both the x and y directions using the IRAF imshift routine, and subtracted it from the unshifted (2007) epoch data. The differences between the narrowband images (2010 minus 2007) are shown in Figure 6, which exhibits how much the shell would have had to have moved in order for us to see any discrepancy between the two epochs. The threshold for seeing movement from 2007 to 2010 lies between a shift of 0.5 and 1 pixels in each of the x and y coordinates. We claim an upper limit to the arc's motion pdot of less than 1 pixel total (0.52 arcsec) in 3 years. 

The angular distance p from Z Cam to the brightest part of the SW arc is 1690 pixels. If the nova ejecta were coasting with no interaction with the ISM then we could place a lower limit on the time t since the last Z Cam eruption of 
t $\ > $ p/pdot = 5 Kyr. At 163 pc, pdot corresponds to an upper limit on the currently observed transverse velocity of the ejecta of v = 138 km/s . This is much slower than the ejection velocities observed for novae, consistent with our conclusion above that the ejecta have evolved into the momentum-conserving snowplow phase of explosions into a uniform medium \citep{cio88}, and have swept up at least several times their original mass of interstellar matter. The time t since the explosion in the snowplow phase is just  r/4v \citep{oor51}. Substituting v $\leq$ 138 km/s and r = 0.7 pc we find t $\geq$ 1300 yr. This makes Z Cam by far the oldest classical nova ever recovered.

\citet{joh07} has suggested that Z Cam was a classical nova observed by Chinese imperial astrologers nearly 21 centuries ago, in October - November 77 BCE \citep{hop62}. Our observations are not inconsistent with this suggestion. If this suggestion is correct then (assuming that the time since eruption is t = r/4v) we predict a current expansion velocity of about 85 km/s, or 0.11 arcsec/year for the Z Cam shell. This should be detectable in just a few years.

\section{Summary and Conclusions}

The current size of the Z Cam ejecta, the measured distance to Z Cam, and the largest observed ejection velocity for novae set a hard lower limit to the time since Z Cam's last nova eruption: 220 years. This places the Z Cam ejecta in the snowplow phase of post-explosion expansion. Images of the brightest part of the nova shell taken in 2007 and 2010 yield an upper limit to the current fractional rate of expansion of the ejecta: rdot/r $\leq$2 x $10^{-4}$. This corresponds to an upper limit to the shell's current transverse velocity of 138 km/s at the 163 parsec distance of Z Cam. These values, in turn, set a lower limit to the age of the last Z Cam nova eruption of 1300 years. Z Cam is thus the oldest classical nova ever recovered. It is the first successful observational test of the prediction, by thermonuclear runaway models of classical novae, that inter-eruption times for novae are in excess of 1,000 years. If Z Cam is the Chinese nova of 77 BCE then we will soon be able to detect its current rate of expansion, which we predict to be about 85 km/s or 0.11 arcsec/yr.

\clearpage

\begin{figure}
\figurenum{1}
\epsscale{1.0}
\plotone{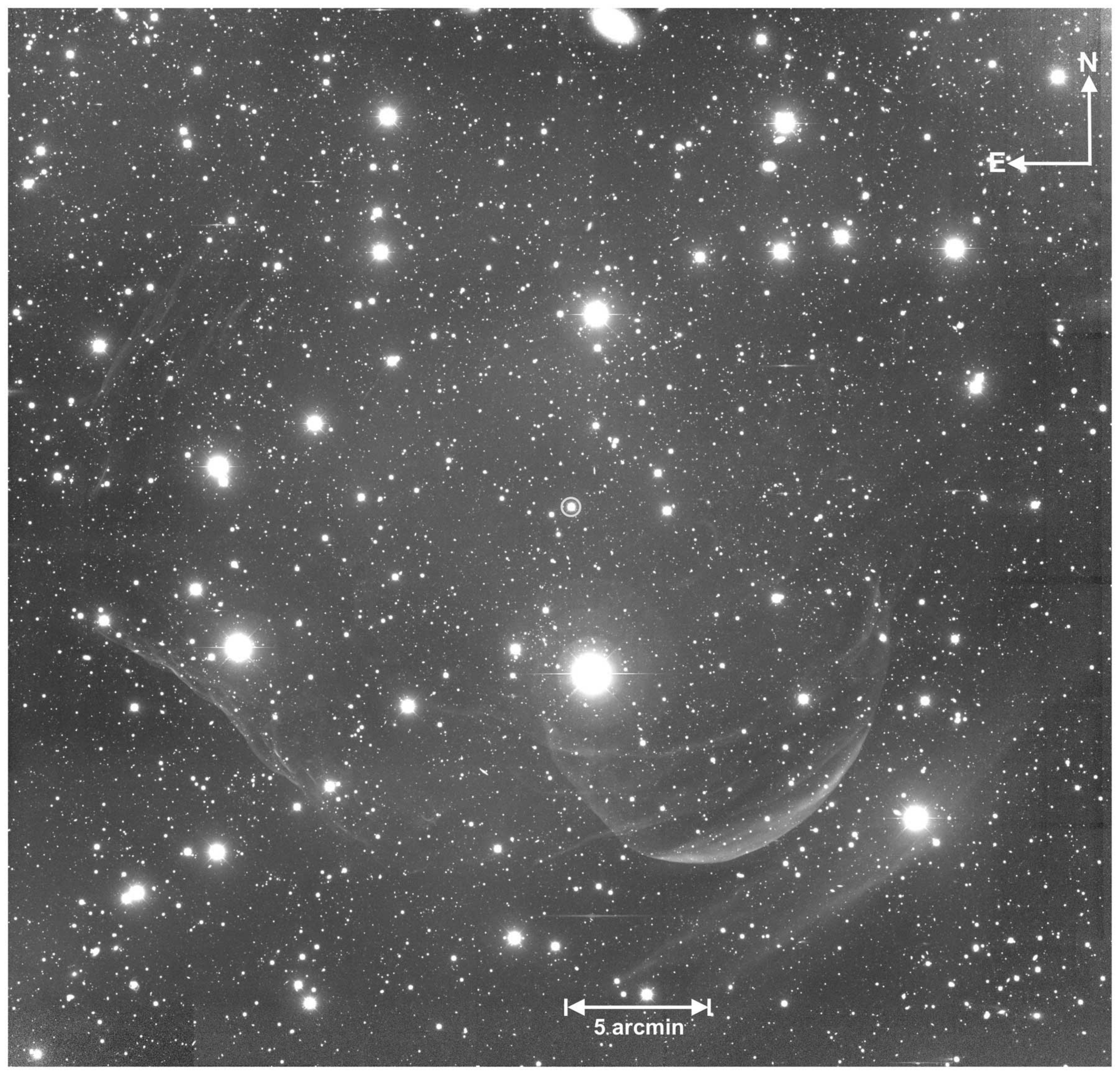}
\caption{An H$\alpha$ +[NII] narrowband image of the dwarf nova Z Camelopardalis and its surroundings. Z Cam is circled.}
\end{figure}

\clearpage

\begin{figure}
\figurenum{2}
\epsscale{1.0}
\plotone{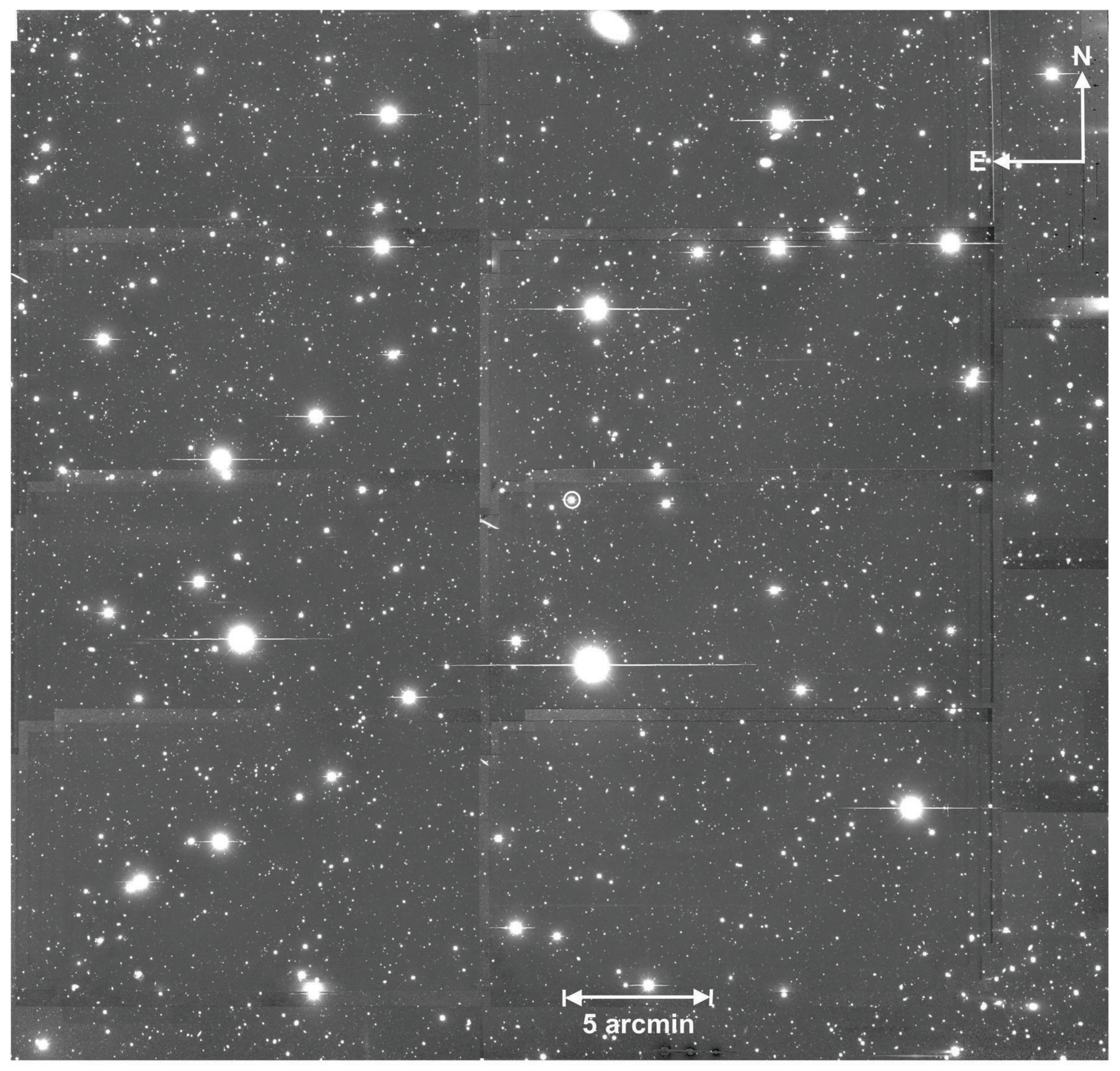}
\caption{A broadband R image of the dwarf nova Z Camelopardalis and its surroundings. Z Cam is circled.}
\end{figure}

\clearpage

\begin{figure}
\figurenum{3}
\epsscale{1.0}
\plotone{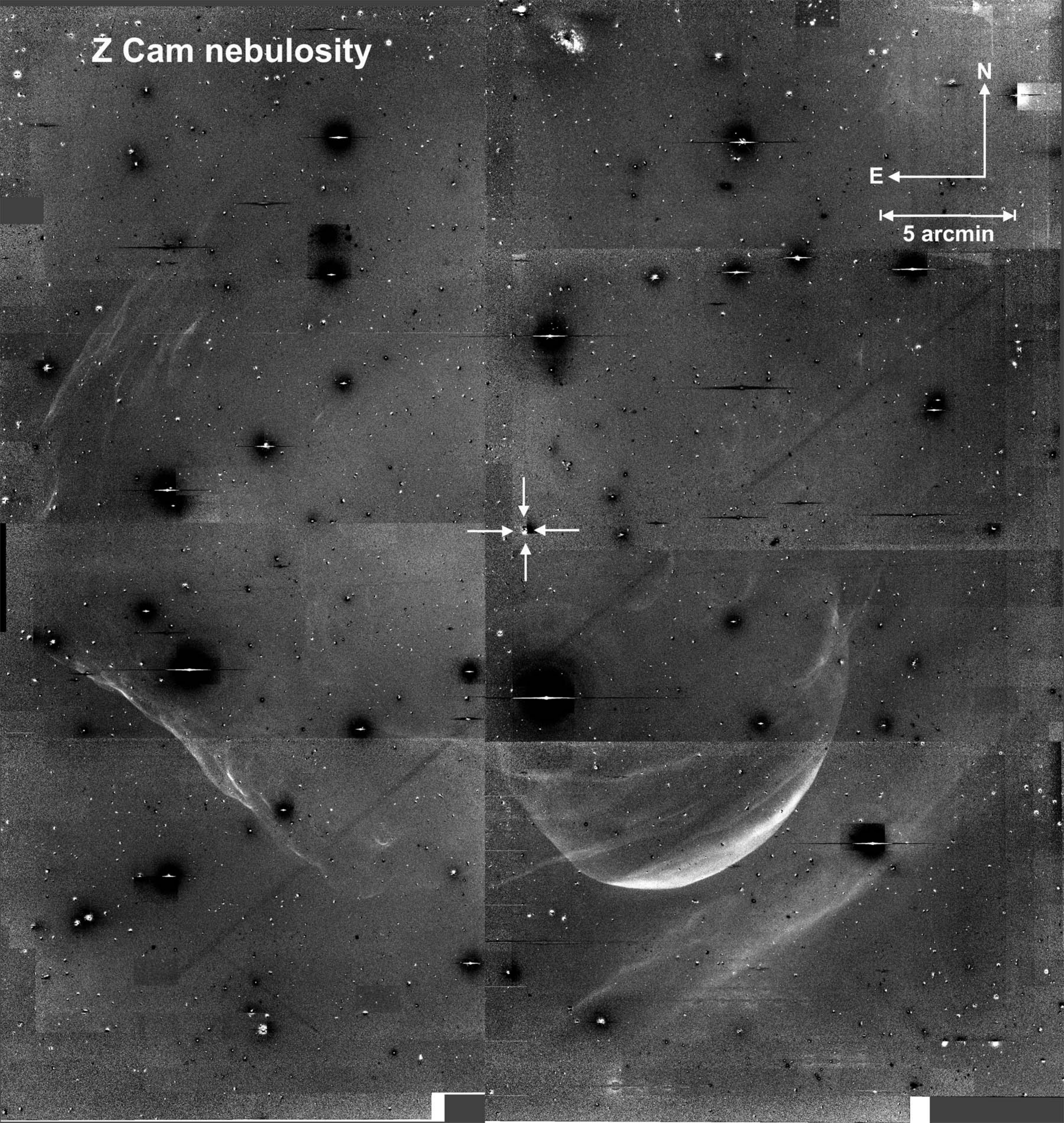}
\caption{A net H$\alpha$ +[NII] image of Z Cam obtained by subtracting Figure 2 from Figure 1. Four arrows point at Z Cam near the center of the image.}
\end{figure}

\clearpage

\begin{figure}
\figurenum{4}
\epsscale{1.0}
\plotone{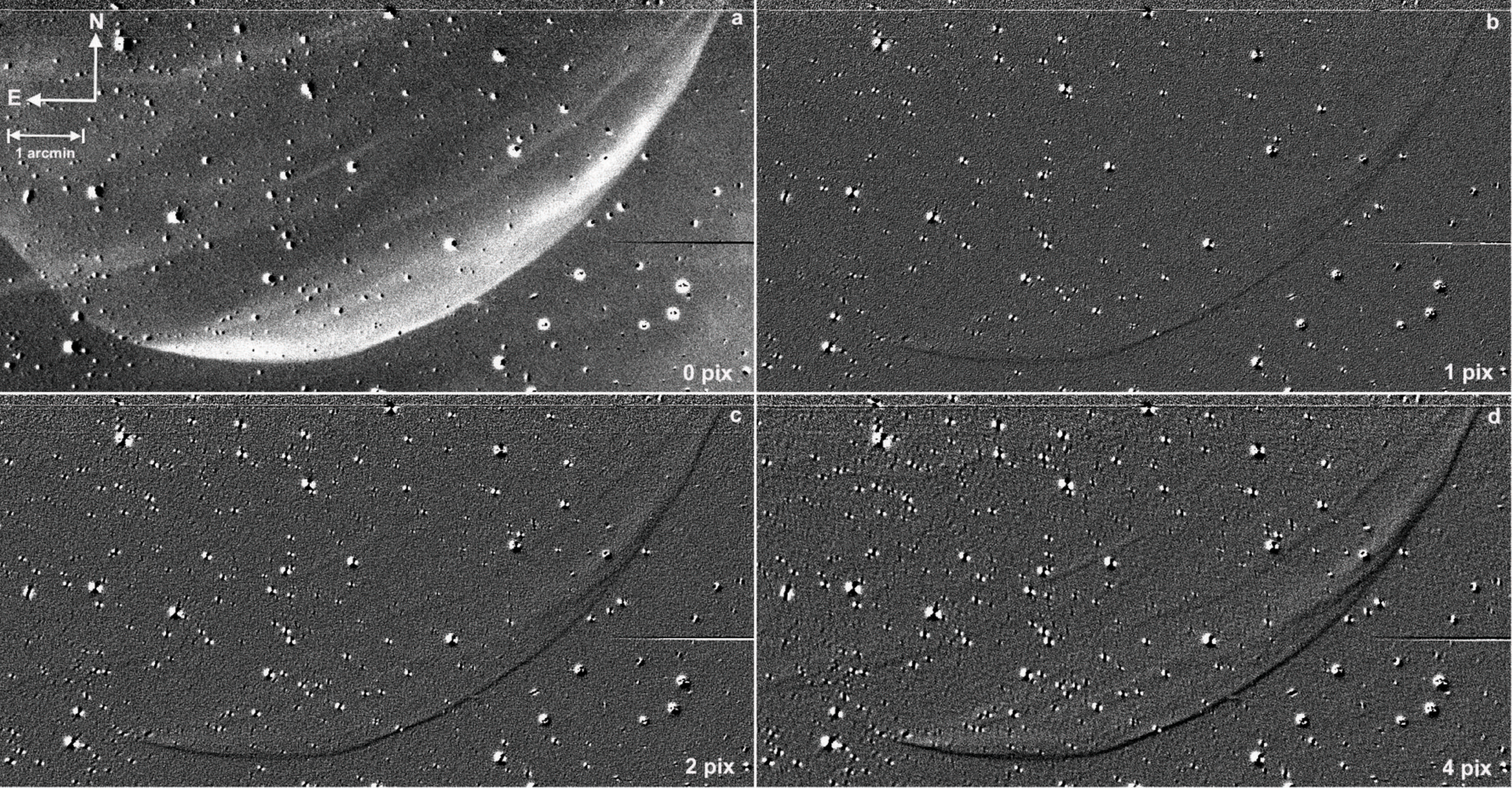}
\caption{Subtle details in the arc of emitting material SW of Z Cam are seen in this collage of images made by offsetting the 2010 image (by the indicated numbers of pixels) from itself in each of the x and y coordinates and then subtracting the displaced image from the original image.  }
\end{figure}

\clearpage

\begin{figure}
\figurenum{5}
\epsscale{1.0}
\plotone{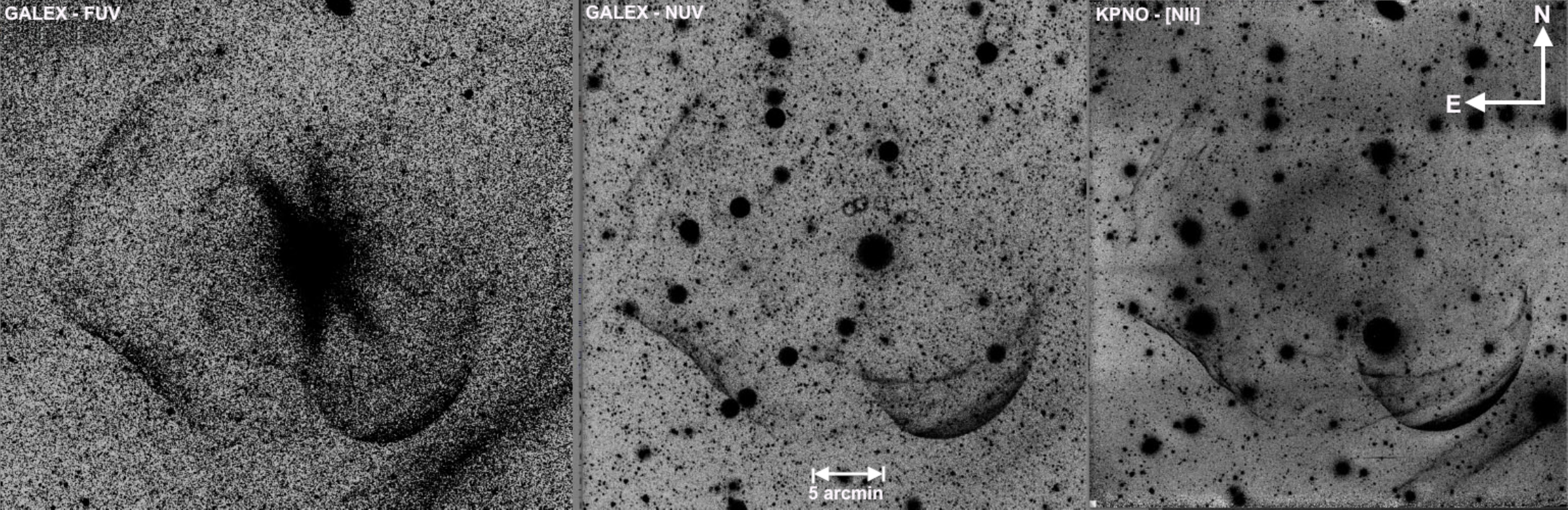}
\caption{A comparison of the GALEX FUV, GALEX NUV and H$\alpha$ +[NII] images of Z Cam and its surroundings.}
\end{figure}

\clearpage

\begin{figure}
\figurenum{6}
\epsscale{1.0}
\plotone{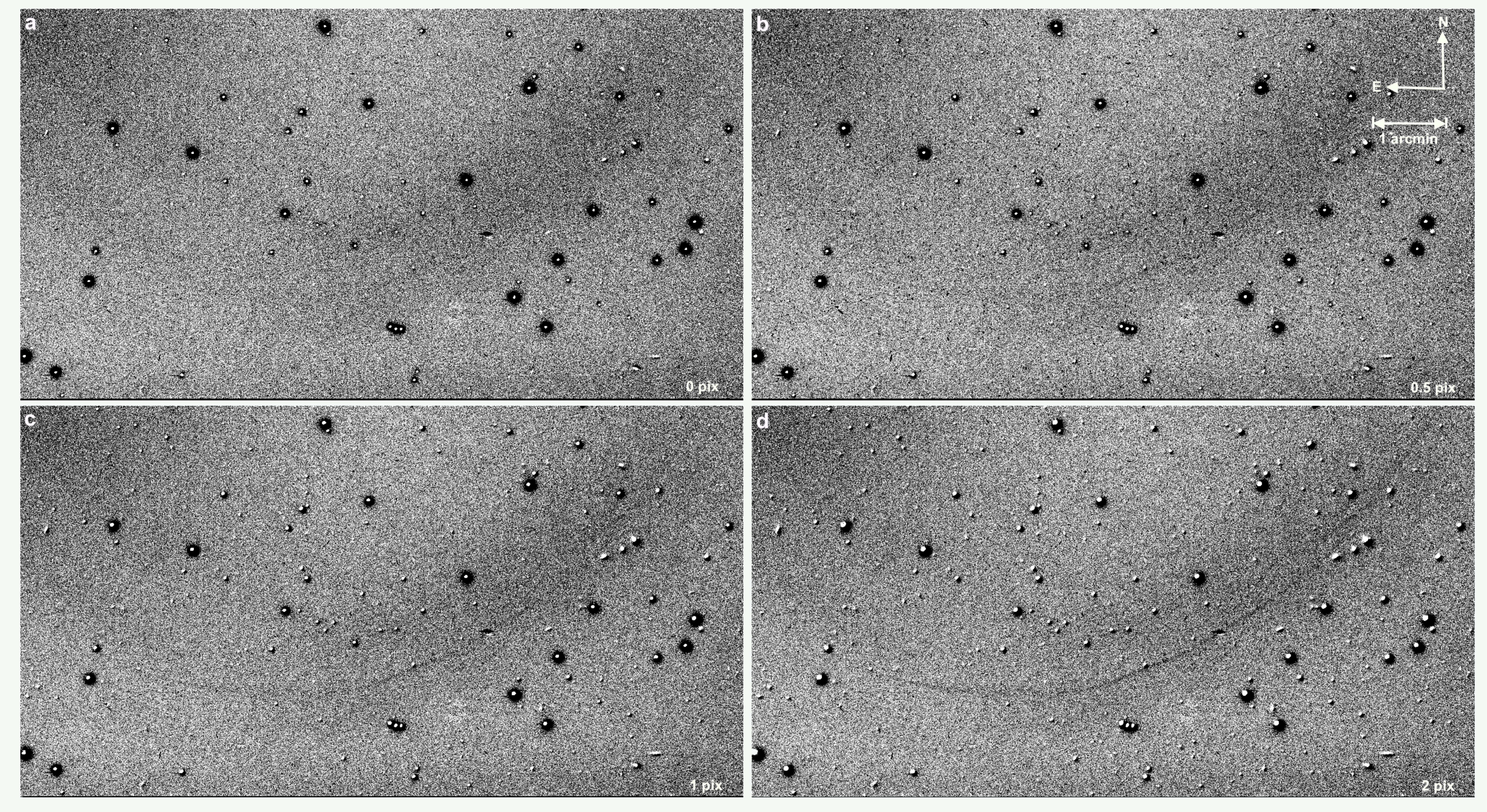}
\caption{The difference between the 2010 and 2007 H$\alpha$ + [NII] images after offsets of 0, 0.5, 1 and 2 pixels in each of x and y are applied. Subtle hints of structure are faintly visible at 0.5 pixels, but we conservatively claim that the upper limit to observed expansion is smaller than 1 pixel (total) during the three years between our observations}.
\end{figure}

\clearpage

\acknowledgments

MMS gratefully acknowledges helpful conversations with, and suggestions from Lars Bildsten, David Kaplan, and Michael Bode.
  
\end{document}